\documentclass[12pt]{iopart}

\expandafter\let\csname equation*\endcsname\relax
\expandafter\let\csname endequation*\endcsname\relax 

\usepackage{amsmath}
\usepackage{iopams,amssymb,epsfig,subfigure,graphicx,amsthm}
\usepackage{amscd}
\usepackage{amsthm}
\usepackage{amssymb}
\usepackage{amsbsy}
\usepackage{makeidx}
\usepackage{epsf}
\usepackage{graphics}
\usepackage{graphicx}
\usepackage{lscape}
\usepackage{amsfonts}
\usepackage{color}
\usepackage{subfigure}

\begin{document}

\title[Variational approximations for travelling solitons]{Variational
approximations for travelling solitons in a discrete nonlinear Schr\"{o}%
dinger equation}
\author{M.\ Syafwan$^{1,2}$, H.\ Susanto$^1$, S.M.\ Cox$^1$ and B.A.\ Malomed%
$^3$}

\address{$^1$ School of Mathematical Sciences, University of Nottingham,\\
University Park, Nottingham NG7 2RD, UK}
\address{$^2$ Department of Mathematics, Faculty of Mathematics and
Natural Sciences,\\
Andalas University, Limau Manis, Padang 25163, Indonesia}
\address{$^3$ Department of Physical Electronics,
School of Electrical Engineering,\\
Tel Aviv University, Tel Aviv 69978, Israel} %
\ead{hadi.susanto@nottingham.ac.uk}

\begin{abstract}
Travelling solitary waves in the one-dimensional discrete nonlinear Schr\"{o}%
dinger equation (DNLSE) with saturable onsite nonlinearity are
studied. A variational approximation (VA) for the solitary waves is
derived in an analytical form. The stability is also studied by
means of the VA, demonstrating that the solitons are stable, which
is consistent with previously published results. Then, the VA is
applied to predict parameters of travelling solitons with
non-oscillatory tails (\textit{embedded solitons}, ESs). Two-soliton
bound states are considered too. The separation distance between the
solitons forming the bound state is derived by means of the VA. A
numerical scheme based on the discretization of the equation in the
moving coordinate frame is derived and implemented using the
Newton--Raphson method. In general, a good agreement between the
analytical and numerical results is obtained. In particular, we
demonstrate the relevance of the analytical prediction of
characteristics of the embedded solitons.
\end{abstract}

\maketitle



\newpage

\section{Introduction}

One of the major issues in studies of spatially discrete systems is whether
such systems can support solitary waves that travel without losing energy to
radiation, which results in deceleration and eventually pinning of the
solitons. The celebrated Peierls-Nabarro (PN) barrier \cite{PN} is the
reason that discrete systems do not generically support exponentially
localized travelling solitary waves. The barrier corresponds to the energy
difference between solutions for the on-site- and the inter-site-centred
lattice solitons, with the latter usually having a higher energy.

In the class of discrete systems, a ubiquitous model with profoundly
important applications in physics and applied mathematics is the discrete
nonlinear Schr\"{o}dinger equation (DNLSE) \cite{Panos}. One- and
two-dimensional (1D and 2D) equations of this type are fundamental models of
discrete nonlinear optics, representing planar and bulk arrays of nonlinear
waveguides coupled by the tunnelling of light between adjacent guiding cores
\cite{discr-optics}. Another well-known application of the DNLSE in any
dimension is the description of Bose--Einstein condensates trapped in deep
optical-lattice potentials, which split the condensate into an array of
droplets \cite{BEC}.

The first attempt at finding travelling lattice solitons in DNLSE was
undertaken in \cite{eilb86}, followed by a more systematic study in \cite%
{fedd91,dunc93}. The latter works indicate that travelling lattice solitons
in the DNLSE are accompanied by nonzero-radiation tails, which was confirmed
by more recent studies \cite{gard04,pelin07,melv09}.

While the most typical onsite nonlinearity in DNLSE models is cubic,
waveguiding arrays made of photorefractive materials feature the saturable
nonlinearity, which strongly facilitates the creation of diverse discrete
solitons \cite{saturable}, including solitary vortices \cite{vortex},
necklace-shaped sets \cite{necklace}, circular solitons \cite{Chen}, and
symmetry-breaking modes \cite{symm-breaking}. Various higher-order soliton
patterns, in the form of stable multi-charged vortices and \textquotedblleft
supervortices" \cite{higher-order} have also been predicted, in the
framework of the same setting. Under appropriate conditions, the saturable
nonlinearity may be approximated by a cubic--quintic truncation; 1D and 2D
solitons in the DNLSE with the cubic--quintic onsite nonlinearity have also
been studied in some detail \cite{CQ}. It was thus found that the saturable
nonlinearity readily supports travelling solitons in discrete media \cite%
{Johansson,Hadzievski}. A reason for this property is the fact that the PN
barrier can change its sign in the case of the saturable nonlinearity \cite%
{Hadzievski}, hence the barrier may vanish at isolated points. This property
may be essential in finding lattice solitary waves that can travel
permanently without emitting radiation (lattice phonons).

The saturable DNLSE modelling the propagation of optical waves in a
photorefractive medium is
\begin{equation}
i\frac{du_{n}}{dt}=-\varepsilon \Delta _{2}u_{n}(t)-\Lambda u_{n}(t)+\frac{%
\sigma u_{n}(t)}{1+|u_{n}(t)|^{2}},  \label{DNLSph}
\end{equation}%
where $u_{n}$ is a complex-valued wave function at site $n$, $\varepsilon $
the strength of the coupling between adjacent sites, $\Delta
_{2}u_{n}(t)=u_{n+1}(t)-2u_{n}(t)+u_{n-1}(t)$ the 1D discrete Laplacian, $%
\Lambda $ a background frequency, and $\sigma $ the nonlinearity
coefficient, which we scale to be $\sigma =+1$, implying that the onsite
nonlinearity is self-focusing. To study travelling-wave solutions of Eq.~(%
\ref{DNLSph}), an ansatz of the form
\begin{equation}
u_{n}(t)=\psi (z,\tau )e^{ikn},  \label{upsi}
\end{equation}%
with $z\equiv n-ct$ and $\tau \equiv t$, is substituted to yield the
time-dependent advance--delay-differential equation
\begin{eqnarray}
&&i\psi _{\tau }(z,\tau )= ic\psi _{z}(z,\tau )+(2\varepsilon -\Lambda )\psi (z,\tau ) \notag \\
&&-\varepsilon \left[ \psi (z+1,\tau
)e^{ik}+\psi (z-1,\tau )e^{-ik}\right] +\frac{\psi (z,\tau )}{1+|\psi
(z,\tau )|^{2}},  \label{decaytau}
\end{eqnarray}%
where $c$ and $k$ are, respectively, the velocity and the wavenumber.

Travelling-wave solutions of Eq.~(\ref{DNLSph}) can be sought using the
time-independent version of Eq.~(\ref{decaytau}),
\begin{equation}
ic\psi ^{\prime }+(2\varepsilon -\Lambda )\psi (z)-\varepsilon \left[ \psi
(z+1)e^{ik}+\psi (z-1)e^{-ik}\right] +\frac{\psi (z)}{1+|\psi (z)|^{2}}=0,
\label{decay}
\end{equation}%
where with $\psi ^{\prime }\equiv \frac{d\psi }{dz}$. Equation~(\ref{decay})
takes a simpler form in the case of $k=0$, following Ref. \cite%
{Melvin2,Melvin}:
\begin{equation}
ic\psi ^{\prime }+(2\varepsilon -\Lambda )\psi (z)-\varepsilon \left[ \psi
(z+1)+\psi (z-1)\right] +\frac{\psi (z)}{1+|\psi (z)|^{2}}=0.
\label{decaytransf}
\end{equation}

The existence of travelling solitons for the DNLSE of this type was
investigated numerically by Melvin \textit{et al}.~\cite{Melvin2,Melvin},
using a pseudo-spectral method to numerically solve Eq.~(\ref{decaytransf}),
which yielded weakly delocalized solitary waves. The delocalization
means that the travelling solitary waves in the saturable DNLSE are, in
general, accompanied by a nonzero oscillating tail, as frequency $\Lambda $
will always resonate with the system's linear spectral (phonon) band.
Because of this, a genuinely travelling solitary wave should be an \textit{%
embedded soliton} (ES), which can exist inside the continuous spectrum, as
an exceptional solution (note that the concept of ESs, which was originally
developed for quiescent solitons in continuous media \cite{embedded}, was
later extended to moving pulses \cite{movingES} and to solitons in dynamical
lattices \cite{discreteES}). Genuinely localized pulse-like solutions were
then generated by finding zeros of the amplitude of the soliton's tail (the
\textquotedblleft tail condition", which is similar to that used to single
out ESs in the continuous family of delocalized intra-band quasi-solitons
\cite{embedded}). The stability of the numerically obtained solutions can
then be analyzed in a numerical form by calculating the Floquet multipliers
of the solutions, using methods similar to that developed in \cite{gard04}.

The presence of genuinely travelling lattice solitons in the saturable DNLSE
in the strong-coupling case has been shown analytically by Oxtoby and
Barashenkov, using exponential asymptotic methods \cite{oxto07} (see also
\cite{melv09}). The use of this sophisticated technique is necessary, as the
radiation emitted by moving solitary waves is exponentially small in the
wave's amplitude. This is a reason why broad, small-amplitude pulses are
highly mobile, seeming like freely travelling solitons.

In this paper we apply, for the first time, a variational approximation (VA)
to the study of travelling solitary waves and their stability, as well as
for predicting the location of the genuinely localized travelling solitary
waves. In the context of DNLSE with cubic nonlinearity, the application of
VA was proposed to construct the fundamental onsite and intersite soliton
solutions by using a trial function containing unknown parameters that have
to be minimized using the Euler-Lagrange equations \cite{Weinst,Kaup}. The
same VA method has been applied recently to the cubic-quintic DNLSE \cite%
{Chong} (see also \cite{Chong11} and references therein). It was shown that
the method is not only excellent in approximating the fundamental discrete
solitons, but also correctly predicts their stability.

The rest of the paper is organized as follows. In Section 2, we develop the
VA for the solitary-wave solutions of the advance-delay equation. In the
same section, we derive an analytical function whose zeros correspond to the
location of ESs. The use of the VA in analyzing the stability of the
travelling solitary waves is then discussed. In addition to single-hump
pulses, in the same section we consider bound states built from two
solitons, and the use of the VA to predict the distance between them. In
Section 3, we introduce a numerical scheme for solving Eq.~(\ref{decaytransf}%
) and compare the numerical results with the analytical calculations
performed in the preceding section. Results of the work are summarized in
Section 4.

\section{The variational approximation}

\subsection{Core soliton solutions}

As suggested by previous work \cite{gard04}, a travelling lattice wave may
be considered as superpositions of an exponentially localized \textit{core}
and extended background built of finite-amplitude plane waves. Here, we
first derive the VA for the core. To this end, we recall that Eq.~(\ref%
{decaytransf}) can be represented in the variational form,
\begin{equation}
\delta L/\delta \psi ^{\ast }(z)=0,  \label{delta}
\end{equation}%
where $\delta /\delta \psi ^{\ast }$ stands for the variational derivative
of a functional, the asterisk denotes complex conjugation, and the
Lagrangian is
\begin{eqnarray}
L &=&\int_{-\infty }^{+\infty }\!\left[ (2\varepsilon -\Lambda )|\psi
|^{2}+\ln \left( 1+|\psi |^{2}\right) +\frac{ic}{2}\left[ \psi ^{\ast }\psi
^{\prime }-\psi (\psi ^{\prime })^{\ast }\right] \right.  \notag \\
&&\left. -\frac{\varepsilon }{2}\left\{ \psi ^{\ast }[\psi (z+1)+\psi
(z-1)]+\psi \lbrack \psi ^{\ast }(z+1)+\psi ^{\ast }(z-1)]\right\} \right]
\,dz.  \label{L}
\end{eqnarray}

A suitable trial function, or \textit{ansatz}, may be chosen as
\begin{equation}
\psi _{\mathrm{core}}(z)=F(z)\exp \left( ipz\right) ,  \label{sech}
\end{equation}%
\begin{equation}
F(z)=A\mathrm{sech}\,(az),  \label{F}
\end{equation}
where $A$, $a$, and $p$ are real variational parameters. While this ansatz
postulates exponential tails of the soliton, the prediction of solitons
within the framework of the VA does not necessarily mean that the
corresponding solitons exist in a rigorous sense, as the actual tail may be
non-vanishing at $|z|\rightarrow \infty $. In fact, the prediction of
solitons by the VA may imply a situation in which the amplitude of the
nonvanishing tail is not zero, but attains its minimum~\cite{Dave}.

The next step is to substitute ansatz (\ref{sech}) into Lagrangian (\ref{L}%
), perform the integration, and derive the Euler-Lagrange equations,
\begin{equation}
\partial L/\partial A=\partial L/\partial a=\partial L/\partial p=0.
\label{var}
\end{equation}%
By substituting ansatz (\ref{sech}) into the Lagrangian and performing the
integration, we obtain the following effective Lagrangian, as a function of
parameters $A$, $a$, and $p$:
\begin{eqnarray}
L_{\mathrm{eff}} &=&\frac{2A^{2}(2\varepsilon -\Lambda -cp)}{a}+\frac{\ln
^{2}\left( \sqrt{1+A^{2}}+A\right) +\ln ^{2}\left( \sqrt{1+A^{2}}-A\right) }{%
a}  \notag \\
&&-\frac{4A^{2}\varepsilon \cos (p)}{\sinh (a)}.  \label{Leff}
\end{eqnarray}%
Then, substituting Lagrangian (\ref{Leff}) into Eqs.~(\ref{var}) yields the
following equations: 
\begin{gather}
\frac{A(2\varepsilon -\Lambda -cp)}{a}+\frac{\ln \left( \sqrt{1+A^{2}}%
+A\right) }{a\sqrt{1+A^{2}}}-\frac{2A\varepsilon \cos (p)}{\sinh (a)}=0,
\label{c1} \\
-\frac{2A^{2}(2\varepsilon -\Lambda -cp)}{a^{2}}-\frac{\ln ^{2}\left( \sqrt{%
1+A^{2}}+A\right) +\ln ^{2}\left( \sqrt{1+A^{2}}-A\right) }{a^{2}}  \notag \\
{}+\frac{4A^{2}\varepsilon \cos (p)\cosh (a)}{\sinh ^{2}(a)}=0,  \label{c2}
\\
-\frac{c}{a}+\frac{2\varepsilon \sin (p)}{\sinh (a)}=0.  \label{c3}
\end{gather}%
%
%
%
%
This system of algebraic equations for $A$, $a$, and $p$ can be solved
numerically.

\subsection{Prediction of the VA for embedded solitons}

We now seek a condition for the possible existence of ESs. To do so, we
begin by considering a delocalized solution of the linearized version of
Eq.~(\ref{decaytransf}), $\psi _{\mathrm{bckg}}(z)$, which represents the
non-vanishing background. Then, following Ref.~\cite{Dave}, it can be shown
that the condition for the possible existence of ESs, i.e., the absence of
nonzero backgrounds attached to the soliton, is the natural orthogonality
relation,
\begin{equation}
\int_{-\infty }^{+\infty }\left\{\left. \delta L/\delta \psi ^{\ast }\right|_{\psi(z)=\psi _{\mathrm{core}}(z)}
\psi _{\mathrm{bckg}}^{\ast }(z)+\mathrm{c.c.}\right\} dz=0,  \label{ortho}
\end{equation}%
where $\mathrm{c.c.}$ stands for the complex conjugate of the preceding
expression and the variational derivative $\left.\delta L/\delta \psi ^{\ast }\right|_{\psi(z)=\psi _{\mathrm{core}}(z)}$ is the left-hand side of Eq.~(\ref{decaytransf}) with $\psi(z)$ replaced by $\psi _{\mathrm{core}}(z)$.
In the context of the VA, $%
\psi_{\mathrm{core}} (z)$ in Eq.~(\ref{ortho}) should be substituted by the (approximate)
form corresponding to the soliton. Here, the background function is taken as
\begin{equation}
\psi _{\mathrm{bckg}}(z)=\psi _{0}\exp \left( i\lambda z\right) ,
\label{tailx}
\end{equation}%
with constant amplitude $\psi _{0}$, while the soliton's core is
approximated by ansatz (\ref{sech}). Next, the frequency $\lambda $ of the
oscillating background in Eq.~(\ref{tailx}) can be found by substituting $%
\psi _{\mathrm{bckg}}$ into the linearization of Eq.~(\ref{decaytransf}),
i.e.,
\begin{equation}
ic\frac{d\psi }{dz}+\left( 2\varepsilon -\Lambda +1\right) \psi
(z)-\varepsilon \left[ \psi (z+1)+\psi (z-1)\right] =0,  \label{lin}
\end{equation}%
which yields
\begin{equation}
c\lambda +(\Lambda -1)-2\varepsilon \left( 1-\cos (\lambda )\right) =0.
\label{sin}
\end{equation}

By setting $\psi _{0}=\alpha +i\beta $, where $\alpha $ and $\beta $ are
non-zero real constants, Eq.~(\ref{ortho}) yields
\begin{equation}
\int_{-\infty }^{+\infty }(\alpha \mathcal{M}+\beta \mathcal{N})dz=0,
\label{kp}
\end{equation}%
where we define functions
\begin{eqnarray}
\mathcal{M}(z) &=&\left\{ \left( 2\varepsilon -c\lambda -\Lambda \right)
F(z)+\frac{F(z)}{1+F(z)^{2}}\right\} \cos ((\lambda -p)z)  \notag \\
&&-\varepsilon \left[ F(z+1)\cos ((\lambda -p)z-p)+F(z-1)\cos ((\lambda
-p)z+p)\right] ,  \label{M} \\
\mathcal{N}(z) &=&\left\{ \left( 2\varepsilon -c\lambda -\Lambda \right)
F(z)+\frac{F(z)}{1+F(z)^{2}}\right\} \sin ((p-\lambda )z)  \notag \\
&&-\varepsilon \left[ F(z+1)\sin ((p-\lambda )z+p)+F(z-1)\sin ((p-\lambda
)z-p)\right] .  \label{N}
\end{eqnarray}%
%
%
%
%
%

It is readily checked that $\mathcal{N}(z)$ is an odd function, while $%
\mathcal{M}(z)$ is an even one. Therefore, after some manipulations,
integral relation~(\ref{kp}) may be cast into the form 
\begin{eqnarray}
\int_{0}^{+\infty }\left[ \displaystyle\frac{(2\varepsilon -c\lambda
-\Lambda )\cos ((\lambda -p)z)}{\cosh (az)}-\displaystyle\frac{2\cosh
(az)\cos ((\lambda -p)z)}{\cosh (2az)+1+2A^{2}}\right. &&  \notag \\
\left. -\displaystyle\frac{\mathcal{B}\varepsilon \cos ((\lambda -p)z)\cosh
(az)}{\cosh (2az)+\cosh (2a)}+\displaystyle\frac{\mathcal{C}\varepsilon \sin
((\lambda -p)z)\sinh (az)}{\cosh (2az)+\cosh (2a)}\right] dz &=&0,
\label{integral1}
\end{eqnarray}%
with $\mathcal{B}\equiv 4\cos (p)\cosh (a)$ and $\mathcal{C}\equiv 4\sin
(p)\sinh (a)$. The integrals in\ the first and the last terms are evaluated
using formulas 3.981-3 and 3.983-5, while the second and the third
terms use 3.984-4, from tables of integrals given in book \cite%
{Gradshteyn}. The calculation yields 

\begin{equation}
E\equiv (2\varepsilon -c\lambda -\Lambda )-2\varepsilon \cos (\lambda )+%
\frac{\cos \left( \left[ (\lambda -p)\cosh ^{-1}(1+2A^{2})\right] /2a\right)
}{\cosh \left( \left[ \cosh ^{-1}(1+2A^{2})\right] /2\right) }=0,
\label{integralfinal}
\end{equation}%
provided that $a>0$ and $\lambda >p$. Thus, in the framework of the VA, Eq.~(%
\ref{integralfinal}), along with the results of the VA for the soliton's
core given by Eqs.~(\ref{c1})--(\ref{c3}), and with Eq.~(\ref{sin}), may
predict a curve---in particular, in the $(\lambda ,\varepsilon )$
plane---along which the existence of the ESs may be expected.

\subsection{The VA-based stability analysis}

Here, we propose to use the VA to study the stability of the core of the
travelling lattice solitary wave by calculating eigenvalues for modes of
small perturbations in the moving coordinate frame, following Ref. \cite%
{flyt93}. The stability of the background, i.e., the\ modulational
(in)stability of the plane lattice waves, was studied in \cite{gard04}.

The underlying time-dependent equation (\ref{decaytau}) with $k=0$
simplifies to
\begin{eqnarray}
&&-ic\psi _{z}(z,\tau )+i\psi _{\tau }(z,\tau )=  \notag \\
&&(2\varepsilon -\Lambda )\psi (z,\tau )-\varepsilon \left[ \psi (z+1,\tau
)+\psi (z-1,\tau )\right] +\frac{\psi (z,\tau )}{1+|\psi (z,\tau )|^{2}}.
\label{decaytau1}
\end{eqnarray}%
The Lagrangian of Eq. (\ref{decaytau1}) is
\begin{eqnarray}
L &=&\int_{-\infty }^{+\infty }\left[ (2\varepsilon -\Lambda )|\psi
|^{2}+\ln \left( 1+|\psi |^{2}\right) +\frac{ic}{2}\left[ \psi ^{\ast }\psi
_{z}-\psi \psi _{z}^{\ast }\right] \right.  \notag \\
&&-\frac{\varepsilon }{2}\left\{ \psi ^{\ast }[\psi (z+1,\tau )+\psi
(z-1,\tau )]+\psi \lbrack \psi ^{\ast }(z+1,\tau )+\psi ^{\ast }(z-1,\tau
)]\right\}  \notag \\
&&\left. -\frac{i}{2}\left( \psi ^{\ast }\psi _{\tau }-\psi \psi _{\tau
}^{\ast }\right) \right] dz.  \label{Lstab}
\end{eqnarray}%
Note that Eq.\ (\ref{decaytau1}) is produced by the variation with respect
to $\psi ^{\ast }$ not of Lagrangian (\ref{Lstab}), but rather of the
corresponding action functional, $S=\int Ldt$. However, for practical
purposes (the derivation of VA equations), it is enough to calculate
Lagrangian (\ref{Lstab}) (it is not necessary to calculate the action
functional explicitly).

The time-dependent ansatz, generalizing the static one (\ref{sech}), is
\begin{eqnarray}
\psi _{\mathrm{core}}(z,\tau ) &=&A(\tau )\mathrm{sech}\left[ a(\tau )(z-\xi
(\tau ))\right]  \notag \\
&&\times \exp \left( i\phi (\tau )+ip(\tau )z+\frac{i}{2}C(\tau )\left[
z-\xi (\tau )\right] ^{2}\right) ,  \label{sechtime}
\end{eqnarray}%
where all parameters are real functions of time. Additional variational
parameters which appear here are the coordinate of soliton's centre, $\xi
(\tau )$, the overall phase, $\phi (\tau )$, and the intrinsic chirp, $%
C(\tau )$. Substituting ansatz (\ref{sechtime}) into Lagrangian (\ref{Lstab}%
) and performing the integration yields the corresponding effective
Lagrangian,
\begin{eqnarray}
L_{\mathrm{eff}} &=&A(\tau )^{2}\left\{ \frac{-2\Lambda +\varepsilon Q(\tau
)+2\left[ \xi (\tau )p^{\prime }(\tau )+\phi ^{\prime }(\tau )-cp(\tau )%
\right] }{a(\tau )}+\frac{\pi ^{2}C^{\prime }(\tau )}{12a(\tau )^{3}}\right\}
\notag \\
&&+\frac{\ln ^{2}\left( \sqrt{1+A(\tau )^{2}}+A(\tau )\right) +\ln
^{2}\left( \sqrt{1+A(\tau )^{2}}-A(\tau )\right) }{a(\tau )},
\label{Leffstab}
\end{eqnarray}%
with primes standing for the derivatives, and%
\begin{eqnarray}
Q(\tau ) &\equiv &4-\displaystyle\frac{4\pi \sin \left( \frac{C(\tau )}{2}%
\right) \cos \left( p(\tau )\right) }{\sinh \left( a(\tau )\right) \sinh
\left( \frac{C(\tau )\pi }{2a(\tau )}\right) }  \notag \\
&=&4-\frac{4a(\tau )\cos (p(\tau ))}{\sinh (a(\tau ))}+\frac{(a(\tau
)^{2}+\pi )\cos (p(\tau ))C(\tau )^{2}}{6a(\tau )\sinh (a(\tau ))}+{\mathcal{%
O}}(C^{4}).
\end{eqnarray}

The Euler--Lagrange equations for the variational parameters take the form
of an ODE system, which may be symbolically written in the vectorial form, $%
\mathbf{\dot{x}}\equiv \left[ A^{\prime }(\tau ),a^{\prime }(\tau
),p^{\prime }(\tau ),C^{\prime }(\tau ),\phi ^{\prime }(\tau ),\xi ^{\prime
}(\tau )\right] ^{T}=\mathbf{g}(\mathbf{x})$, and solved numerically. The
VA-predicted stability analysis is based on the linearization,
\begin{equation}
\mathbf{z}=\mathbf{x}_{0}+\delta \mathbf{y},  \label{ansatzEVP}
\end{equation}%
with infinitesimal $\delta $, and $\mathbf{x}_{0}=\left[ A,a,p,C=0,\phi
=0,\xi =0\right] ^{T}$ representing solutions of static variational
equations\ (\ref{c1})--(\ref{c3}). The substitution of this into the
dynamical Euler--Lagrange equations and linearization leads to an eigenvalue
problem,
\begin{equation}
\mathbf{\dot{y}}=\mathbf{H}\mathbf{y},  \label{evpVA}
\end{equation}%
with the corresponding stability matrix $\mathbf{H}$. The stability of the
stationary solution is then determined by the eigenvalues $\Omega $ of Eq.~(\ref%
{evpVA}), which must be found in a numerical form, the solution being stable
if $\text{Re}(\Omega )\leq 0$ for all eigenvalues.

\subsection{The effective potential of the soliton--soliton interaction and
the formation of bound states}

DNLSEs are known to admit bound states of fundamental (single-humped)
solitons \cite{kivs98,kapi01,kevr01}. The present model also supports bound
states, in addition to the single-hump solitons. In the infinite domain,
there are infinitely many different bound states. An essential feature of
such states is that the distances between the bound solitons are not
arbitrary. In the weak-interaction limit, this feature may be explained by
means of the VA, as we indicate below.

The effective potential for the interaction between two \emph{identical}
solitons separated by distance $|l|$, which is essentially larger than the
width of each soliton, and with a phase shift $\phi $ between them ($\phi
\equiv \phi _{2}-\phi _{1}$, where $\phi _{1,2}$ are the phases at central
points of the two solitons), can be derived following the general approach
elaborated in Refs.~\cite{general,general2}. To this end, we consider the
wave field in the vicinity of one of the two solitons (say, soliton No.~1),
whose centre is located at $z=0$, while the other soliton (No.~2) is located
far afield, at $z=l$; thus we set
\begin{equation}
\psi (z)=\psi _{1}(z)+\psi _{2}(z).  \label{psi}
\end{equation}%
Here $\psi _{2}$ is realized as a weak tail of the second soliton (of
course, the tail is affected by the overlap with soliton No.~1). Then,
expression (\ref{psi}) is substituted into the Hamiltonian ($H$)
corresponding to Lagrangian (\ref{L}). The effective interaction potential
is represented by the corresponding term in the total Hamiltonian which is
linearized in the weak field, $\psi _{2}$. Thus, the corresponding
contribution to the potential is
\begin{equation}
U_{12}=\int_{-\infty }^{+\infty }\left[ \frac{\delta H}{\delta \psi (z)}\psi
_{2}(z)+\frac{\delta H}{\delta \psi ^{\ast }(z)}\psi _{2}^{\ast }(z)\right]
dz,  \label{12}
\end{equation}%
where the variational derivatives are taken at $\psi _{2}=0$ (i.e., for $%
\psi =\psi _{1}$, see Eq.~(\ref{psi})). The integral is formally written
over an infinite domain, although it is assumed that it will be calculated
in a vicinity of the first soliton (see below).

Because the stationary one-soliton solution is itself found from the equation
\begin{equation}
\frac{\delta H}{\delta \psi (z)}=0,  \label{zero}
\end{equation}%
with $\psi =\psi _{1}$, it might seem that expression (\ref{12}) should
identically vanish (the variational derivatives corresponding to the
single-soliton solution should be zero, according to Eq.~(\ref{zero})).
However, the integral will in fact vanish only after performing the
integration by parts of the terms which contain the first derivative of $%
\psi _{2}$:
\begin{equation}
\frac{ic}{2}\int_{-\infty }^{+\infty }\left[ \psi _{1}^{\ast }\frac{\partial
\psi _{2}}{\partial z}-\psi _{1}\left( \frac{\partial \psi _{2}}{\partial z}%
\right) ^{\ast }\right] dz.  \label{deriv}
\end{equation}%
Thus, the sole nonzero contribution to $U_{12}$ originates from the surface
term produced by the integration by parts in expression (\ref{deriv}): $%
U_{12}=\left( ic/2\right) \Delta \left\{ \psi _{1}^{\ast }\psi _{2}-\psi
_{1}\psi _{2}^{\ast }\right\} ,$ where the notation $\Delta \left\{
...\right\} $ denotes the difference between the values of this expression
at two arbitrary points, $Z_{-}$ and $Z_{+}$, located sufficiently far from
the centre of the first soliton (on its left and right sides, respectively),
but so that the second soliton remains much further still. In fact, one can
take $Z_{-}=-\infty $, while $Z_{+}$ is an arbitrary intermediate point
between the two solitons, located far from both, but closer to the first
soliton. Finally, the total interaction potential also includes a symmetric
contribution from the vicinity of the second soliton:
\begin{equation}
U_{\mathrm{int}}=U_{12}+U_{21}=\frac{1}{2}ic\Delta \left( \left\{ \psi
_{1}^{\ast }\psi _{2}-\psi _{1}\psi _{2}^{\ast }\right\} +\left\{ \psi
_{2}^{\ast }\psi _{1}-\psi _{2}\psi _{1}^{\ast }\right\} \right) .  \label{U}
\end{equation}

The main trick (which was employed in Refs.~\cite{general,general2}) is to
use the asymptotic expressions for the wave fields of both solitons taken
far from their respective centres (i.e., their \emph{tails}). To this end,
we first consider the \emph{linearized} equation (\ref{lin}), with soliton
tails sought in the form of
\begin{equation}
\psi (z)=\psi ^{(0)}e^{\lambda _{\mathrm{ap}}z},  \label{psi0}
\end{equation}%
where constant $\psi ^{(0)}$ is determined by the full nonlinear solution.
In fact, expression (\ref{psi0}) can be viewed as a limiting form of ansatz (%
\ref{sech}) as $|z|\rightarrow \infty $, such that $\psi ^{(0)}=2A$ and $%
\lambda _{\mathrm{ap}}=a+ip$. Note the similarity between expressions (\ref%
{psi0}) and (\ref{tailx}), the only difference being that $\lambda $ was
real, whereas $\lambda _{\mathrm{ap}}$ may be complex. Replacing $\lambda $
by $-i\lambda _{\mathrm{ap}}$ in (\ref{sin}), we find that $\lambda _{%
\mathrm{ap}}$ then satisfies equation%
\begin{equation}
ic\lambda _{\mathrm{ap}}+\left( 2\varepsilon -\Lambda +1\right)
-2\varepsilon \cosh (\lambda _{\mathrm{ap}})=0.  \label{cosh}
\end{equation}%
It is evident that if $\lambda _{\mathrm{ap}}$ is a complex root of Eq.~(\ref%
{cosh}), then $-\lambda _{\mathrm{ap}}^{\ast }$ is also a root. Thus, the
pair of the complex roots may be defined through their real and imaginary
parts as $\pm \lambda _{r}+i\lambda _{i}$, where $\lambda _{r}$ is chosen to
be positive, by definition. In this regard, the transcendental complex
equation (\ref{cosh}) may be cast into an explicit form if $\lambda _{r}$
and $\lambda _{i}$ are treated as free parameters, while $c$ and $\Lambda $
are considered as unknowns. This approach yields
\begin{eqnarray}
c &=&\frac{2\varepsilon }{\lambda _{r}}\sinh \left( \lambda _{r}\right) \sin
\left( \lambda _{i}\right) ,  \label{c} \\
\Lambda &=&1+2\varepsilon \left[ 1-\cosh \left( \lambda _{r}\right) \cos
\left( \lambda _{i}\right) +\frac{\lambda _{i}}{\lambda _{r}}\sinh \left(
\lambda _{r}\right) \sin \left( \lambda _{i}\right) \right] .  \label{Lambda}
\end{eqnarray}%
Then Eq.~(\ref{psi0}) yields the tails of the two solitons in the form of
\begin{eqnarray}
\psi _{1}(z) &\approx &\psi ^{(0)}\exp \left( -\lambda _{r}|z|+i\lambda
_{i}z\right) ,  \label{bckg1} \\
\psi _{2}(z) &\approx &\psi ^{(0)}\exp \left( -\lambda _{r}|z-l|+i\lambda
_{i}\left( z-l\right) \right)  \label{bckg2}
\end{eqnarray}%
(recall that the centres of solitons No.~1 and~2 are assumed to be located
at $z=0$ and $z=l$, respectively). The substitution of expression (\ref%
{bckg1})--(\ref{bckg2}) into Eq.~(\ref{U}) yields an explicit result, which
\emph{does not }depend on arbitrary intermediate point $Z_{+}$ appearing in
the expression for $U_{12}$ (nor does it depend on the counterpart of $Z_{+}$
arising in $U_{21}$), because contributions from $Z_{+}$ \emph{cancel out}
in the final expression (cf.~Refs.~\cite{general,general2}). We thus obtain
the following expression for the interaction potential (\ref{U}):
\begin{equation}
U_{\mathrm{int}}(l)=2c\left\vert \psi ^{(0)}\right\vert ^{2}\exp \left(
-\lambda _{r}l\right) \sin \left( \lambda _{i}l-\phi \right) ,  \label{int}
\end{equation}%
where the above-mentioned phase shift is taken into account. In this
expression, everything is known, in principle (recall that $\psi ^{(0)}$ is
to be found from the full nonlinear solution for one soliton), if phase
shift $\phi $ is considered as a given \emph{frozen} constant.

Strictly speaking, the exponentially decaying potential (\ref{int}) is valid
for solitons whose waveforms are fully localized. If the actual shape of the
solitons features small nonvanishing oscillating tails, the asymptotic form
of the potential at large values of $l$ will change accordingly.

It is straightforward to see that potential (\ref{int}) gives rise to a set
of local minima and maxima (as a function of $l$), which may correspond to a
series of two-soliton bound states, as well as to more complex multi-soliton
bound states. The extrema of the potential are located at points
\begin{equation}
l_{n}=\frac{1}{\lambda _{i}}\arctan \left( \frac{\lambda _{i}}{\lambda _{r}}%
\right) +\frac{\phi +\pi n}{\lambda _{i}}~,~n\cdot \mathrm{sign}\left(
\lambda _{i}\right) =0,1,2,3,\dots ~.  \label{ln}
\end{equation}%
The extrema are potential minima for even or odd integers (i.e., for $n\cdot
\mathrm{\ sign}\left( \lambda _{i}\right) =0,2,4,\dots $ or $n\cdot \mathrm{%
sign}\left( \lambda _{i}\right) =1,3,5,\dots $, severally), at
$c\lambda _{i}/\lambda _{r}<0$ and
$c\lambda _{i}/\lambda _{r}>0$, respectively. Note that the separation
between the potential minima, $\Delta l=\pi /\left\vert \lambda
_{i}\right\vert $, does not depend on the frozen phase shift, $\phi $.

\section{Numerical scheme and comparisons with analytical results}

\label{findiff}

To solve Eq.~(\ref{decaytransf}) numerically, we use a scheme based on the
discretization of the equation, resulting in a system of difference
equations. We employ central finite differences, so that the corresponding
Jacobian matrix is sparse. The difference equations are then solved using
the Newton--Raphson method. This is different from the previously used
pseudo-spectral collocation method \cite{Melvin}, in which the dependent
variable $\psi $ was represented as a Fourier series, whose coefficients
were then determined by solving a system of algebraic equations obtained by
requiring the series approximation to satisfy the governing equation at
collocation points.

In the framework of the finite-difference method, with grid size $\Delta z$%
,we approximate $\psi (z)$ on a finite interval, $[-L,+L]$, as follows: $%
\psi (z)\rightarrow \psi (n\Delta z)\equiv \psi _{n}$, $\psi (z\pm
1)\rightarrow \psi ((n\pm 1/\Delta z)\Delta z)=\psi _{n\pm 1/\Delta z}$. For
$\psi ^{\prime }(z)\equiv \frac{d\psi }{dz}$, we use either the central
two-point stencil, $(\psi _{n+1}-\psi _{n-1})/(2\Delta z)$, or the spectral
collocation method \cite{weid00,tref00}. In the following, we describe
details of the numerical scheme for the two-point stencil, although the
spectral collocation method has been implemented too.

Substituting the above discretizations into Eq.~(\ref{decaytransf}) yields
\begin{equation}
\frac{ic}{2\Delta z}(\psi _{n+1}-\psi _{n-1})+(2\varepsilon -\Lambda )\psi
_{n}-\varepsilon (\psi _{n+1/\Delta z}+\psi _{n-1/\Delta z})+\frac{\psi _{n}%
}{1+|\psi _{n}|^{2}}=0.  \label{decay1}
\end{equation}%
The number of grid points is $N=2L/\Delta z +1$. We use periodic boundary
conditions, so that $\psi _{N-1+j}=\psi _{j}$ and $\psi _{1-j}=\psi _{N-j}$
for $j=1,2,\dots ,M$, where $M\equiv 1/\Delta z$.

Next, we solve the resulting system of nonlinear equations (\ref{decay1})
numerically for a fixed set of parameters $(c,\varepsilon ,\Lambda )$, using
the Newton--Raphson method with an error tolerance of order $10^{-15}$. To
do so, we define the left-hand side of Eq.~(\ref{decay1}) as $f_{n}$
and then seek a solution with $f_{n}=0$ for $n=1,2,...,N-1$. Because $\psi $ is
complex, we seek solutions in the form of $\psi =\text{Re}(\psi )+i\text{%
Im}(\psi )$. Accordingly, we define a (real) functional vector, $\mathbf{F}%
\equiv \lbrack \text{Re}(f_{1}),\dots ,\text{Re}(f_{N-1}),\text{Im}%
(f_{1}),\dots ,\text{Im}(f_{N-1})]^{T}$, and a (real) solution vector, $%
\mathbf{\Psi }\equiv \lbrack \text{Re}(\psi _{1}),\dots ,\text{Re}(\psi
_{N-1}),\text{Im}(\psi _{1}),\dots ,\text{Im}(\psi _{N-1})]^{T}$. Note that Eq.~(%
\ref{decaytransf}) has rotational and translational invariance. Therefore,
to ensure the uniqueness of solutions, we impose two constraints,
\begin{eqnarray}
\left. 
\begin{array}{ccl}
f_{\frac{N+1}{2}+1} &=&\text{Re}\left( \psi _{\frac{N+1}{2}-1}\right) -\text{%
Re}\left( \psi _{\frac{N+1}{2}+1}\right) =0,  \\
f_{\frac{N+1}{2}+N} &=&\text{Im}\left( \psi _{\frac{N+1}{2}}\right) =0.
\end{array}
\right.
\label{constraints}
\end{eqnarray}%
These constraints significantly improve the convergence of the
Newton--Raphson scheme.

Strictly speaking, we do not have a rigorous proof of the convergence of the
\ numerical scheme outlined above. Therefore, to check the validity of our
findings, we benchmarked the results against those reported in \cite{Melvin}
for the same parameter values. The outputs (not shown here) were indeed
identical for $\Delta z$ small enough and $L$ large enough. Below, we
display results for $L=50$ and $\Delta z=0.2$, which is sufficient for 
clear presentation. We have also used smaller values of $\Delta z$ and
larger $L$, which confirmed the robustness of the results.

\subsection{The soliton's core}

\begin{figure}[tbph]
\centering
\subfigure[$\varepsilon=1$, $c=0.7$, and $\Lambda=0.5$] { \label{comparedprofile1}
  \includegraphics[width=7cm,clip=]{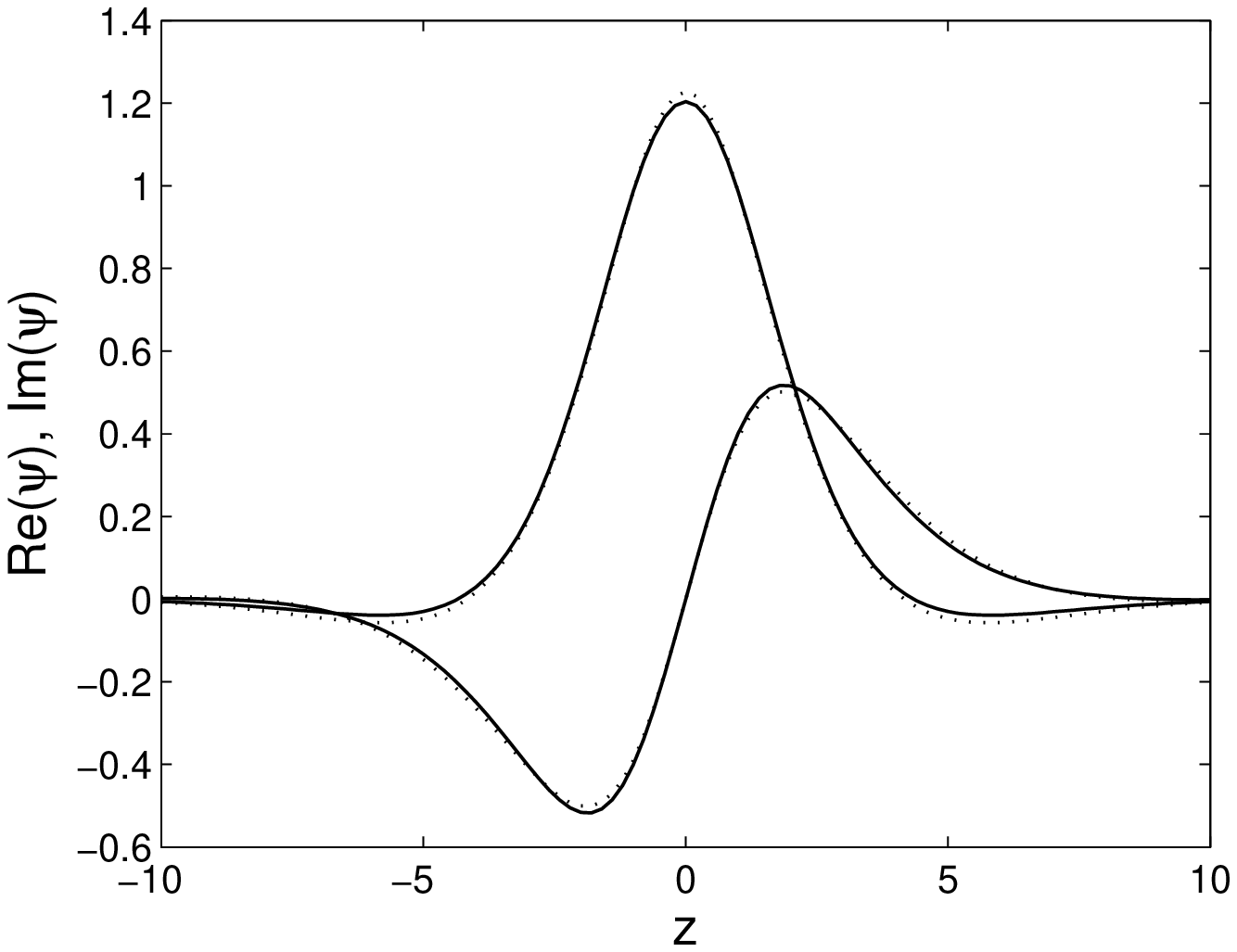}
  } \hspace{0cm}
\subfigure[$\varepsilon=1$, $c=0.7$, and $\Lambda=0.7$] { \label{comparedprofile2}
  \includegraphics[width=7cm,clip=]{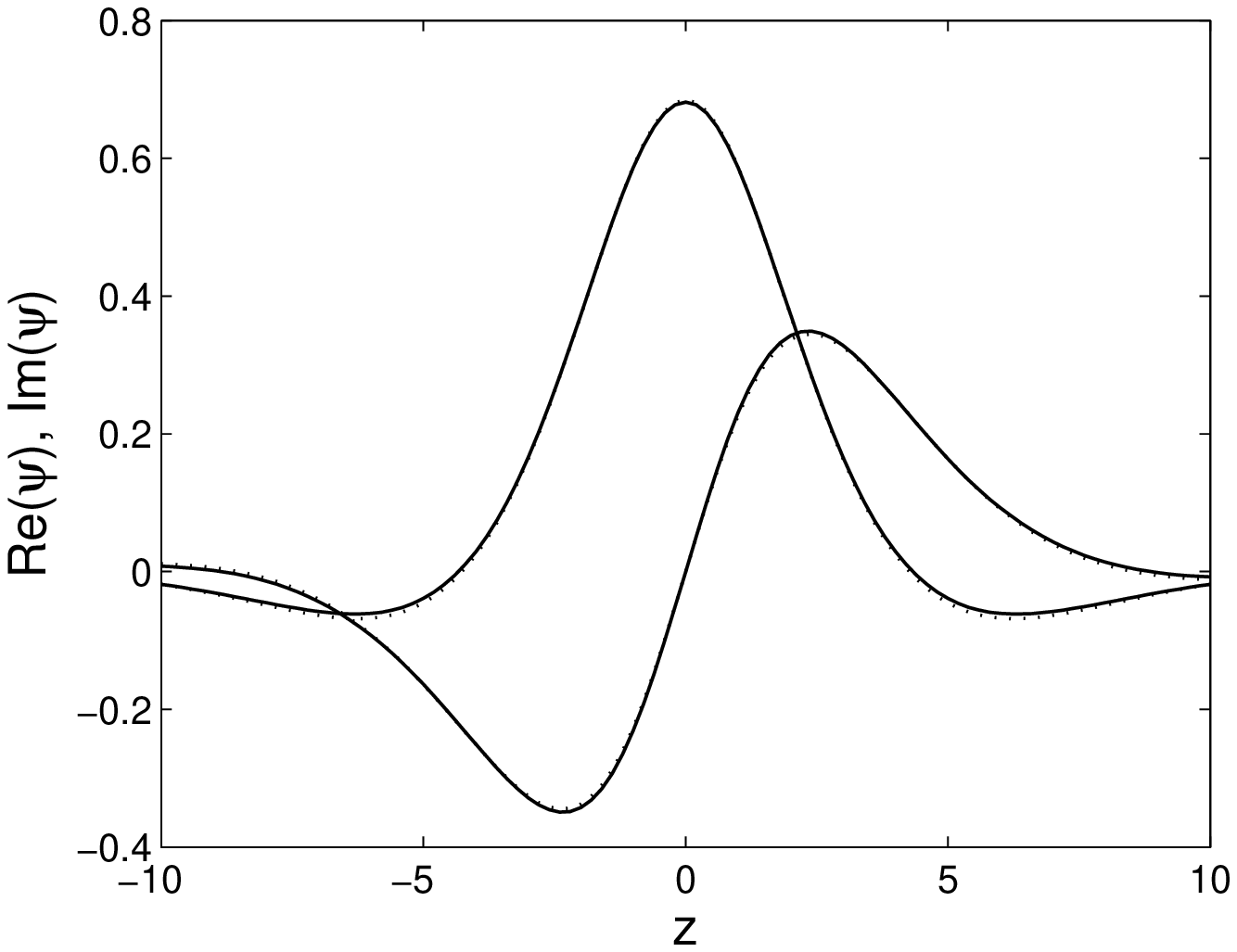}
  }
\caption{The comparison of two soliton profiles for two different values of $%
\Lambda $, as indicated in the caption to each panel. The solid lines
correspond to numerical results, i.e., solutions of Eq.~(\ref{decay1}) imposed with constraints (\ref{constraints}), with $\Delta z=0.2$ and $L=50$ (the $z$-axis
is truncated to focus the picture on the soliton's core), both the real and
imaginary parts being shown. The dotted lines are predictions of the
variational approximation obtained through solving Eqs.~(\ref{c1})--(\ref{c3}).}
\label{comparedprofile}
\end{figure}

For given parameters $\varepsilon ,c$, and $\Lambda $, we solved variational
equations (\ref{c1})--(\ref{c3}) to produce suitable real solutions $A,a$
and $p$, which then give a quasi-analytical approximation for the soliton's
core described by function $\psi _{\mathrm{core}}(z)$ (see Eq.~(\ref{sech}%
)). As generic examples, in Fig.~\ref{comparedprofile} we present, at $%
\varepsilon =1$ and $c=0.7$, the comparison of two soliton profiles obtained
from the numerical results and VA for two different values of $\Lambda $. We
have found $A\approx 1.228,\,a\approx 0.554$, $p\approx 0.377$ for $\Lambda
=0.5$ (Fig.~\ref{comparedprofile1}), and $A\approx 0.685,\,a\approx 0.414,$ $%
p\approx 0.368$ for $\Lambda =0.7$ (Fig.~\ref{comparedprofile2}). Particularly
good agreement is observed in both cases. 


To further confirm the agreement for different values of $\Lambda $, $%
\varepsilon $ and $c$, in the following we compare parameters $A,a$ and $p$
produced by the VA with their numerical counterparts. To calculate the
latter, we make use of the following relations, which are generated by
ansatz (\ref{sech}), (\ref{F}) at $z=0$:
\begin{equation}
\psi (0)=A,\,\psi ^{\prime }(0)=ipA,\,\psi ^{\prime \prime
}(0)=-A(a^{2}+p^{2}),
\label{AapVA}
\end{equation}%
and take the left-hand sides of these relations from numerical data. Thus,
using the central finite differences, we obtain the numerical counterparts
of $A$, $p$, and $a$:
\begin{eqnarray}
A_{\mathrm{num}} &=&\psi _{\frac{N+1}{2}},  \label{Anum} \\
p_{\mathrm{num}} &=&\frac{\mathrm{Im}\left( \psi _{\frac{N+1}{2}+1}\right) -%
\mathrm{Im}\left( \psi _{\frac{N+1}{2}-1}\right) }{2A\Delta z}, \label{pnum} \\
a_{\mathrm{num}} &=&\pm \sqrt{-\frac{\mathrm{Re}\left( \psi _{\frac{N+1}{2}%
-1}\right) -2\mathrm{Re}\left( \psi _{\frac{N+1}{2}}\right) +\mathrm{Re}%
\left( \psi _{\frac{N+1}{2}+1}\right) }{A_{\mathrm{num}}(\Delta z)^{2}}-p_{%
\mathrm{num}}^{2}}. \label{anum}
\end{eqnarray}%
%
%
%

\begin{figure}[tbph]
\centering
\subfigure[] { \label{AapVAxNum1}
  \includegraphics[width=13cm,clip=]{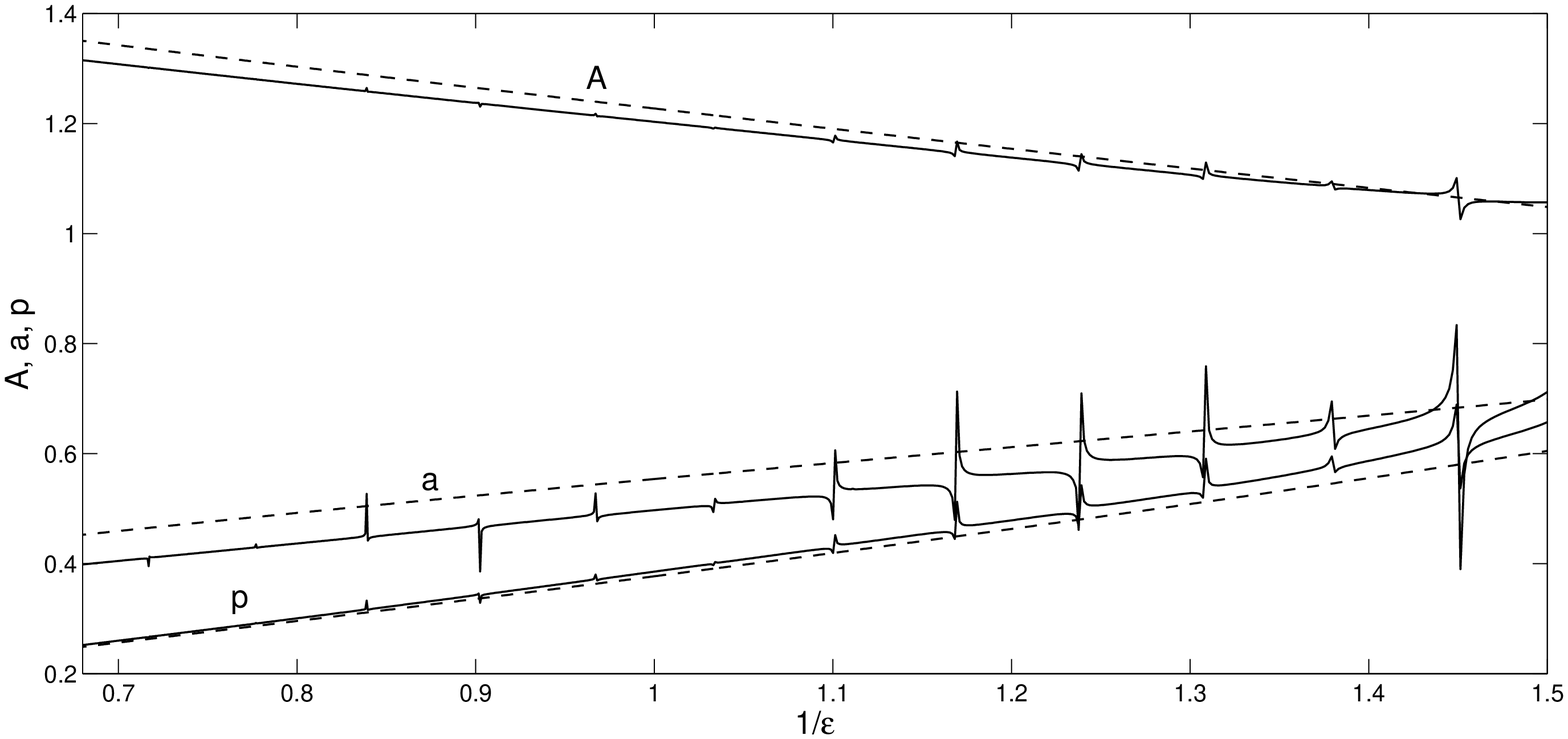}
  } 
\centering
\subfigure[] { \label{AapVAxNum2}
  \includegraphics[width=13cm,clip=]{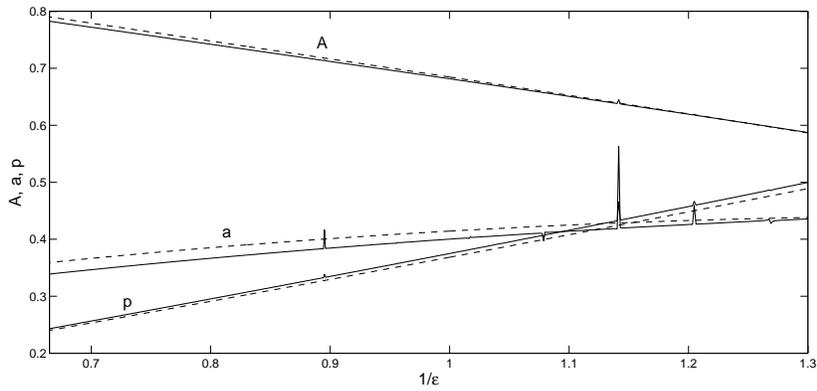}
  }
\caption{The comparison of parameters $(A,a,p)$ calculated from the
numerical data (solid lines) through Eqs.~(\ref{Anum})--(\ref{anum}), and from the VA (dashed lines) through Eqs.~(\ref{AapVA}), for $c=0.7$ and $\Lambda =0.5$ (a) or $\Lambda =0.7$ (b) for varying $1/\protect%
\varepsilon $. }
\label{Aap}
\end{figure}

The comparison of parameters $(A,a,p)$ obtained numerically (solid line) and
from the VA (dashed line) is shown in Fig.~\ref{AapVAxNum1} for varying $%
1/\varepsilon $ and fixed $(c,\Lambda )=(0.7,0.5)$.
We observe that the solid and dashed curves are generally close for all the
three parameters. Nevertheless, we also obtain isolated values of $%
1/\varepsilon $, which behave as singular points. Near the singularities the
numerical results deviate very rapidly from the predictions of the VA. In
fact, at these singular points the numerically obtained solutions are
strongly delocalized due to the resonance of 
the oscillating tails with the finite size of the computational domains,
hence the positions of such singularities depend on $L$, and they may be
considered as artifacts of approximating the infinite region by the finite
domain.


Similarly to Fig.\ \ref{comparedprofile}, where a better approximation is
obtained for larger $\Lambda $, we also observe in Fig.~\ref{AapVAxNum2}
that the variational and numerical curves for $A,\,a,\,p$ are closer for $%
\Lambda =0.7$ than those in \ref{AapVAxNum1}.
In the latter case, the singularities are present too, even though they are
less pronounced here.

Thus we can conclude that the VA provides a reliable approximation for
description of the soliton's core.%



\subsection{Embedded solitons}

\begin{figure}[tbph]
\centering
\subfigure[] { \label{zerocrossmeasure1}
\includegraphics[width=13cm,clip=]{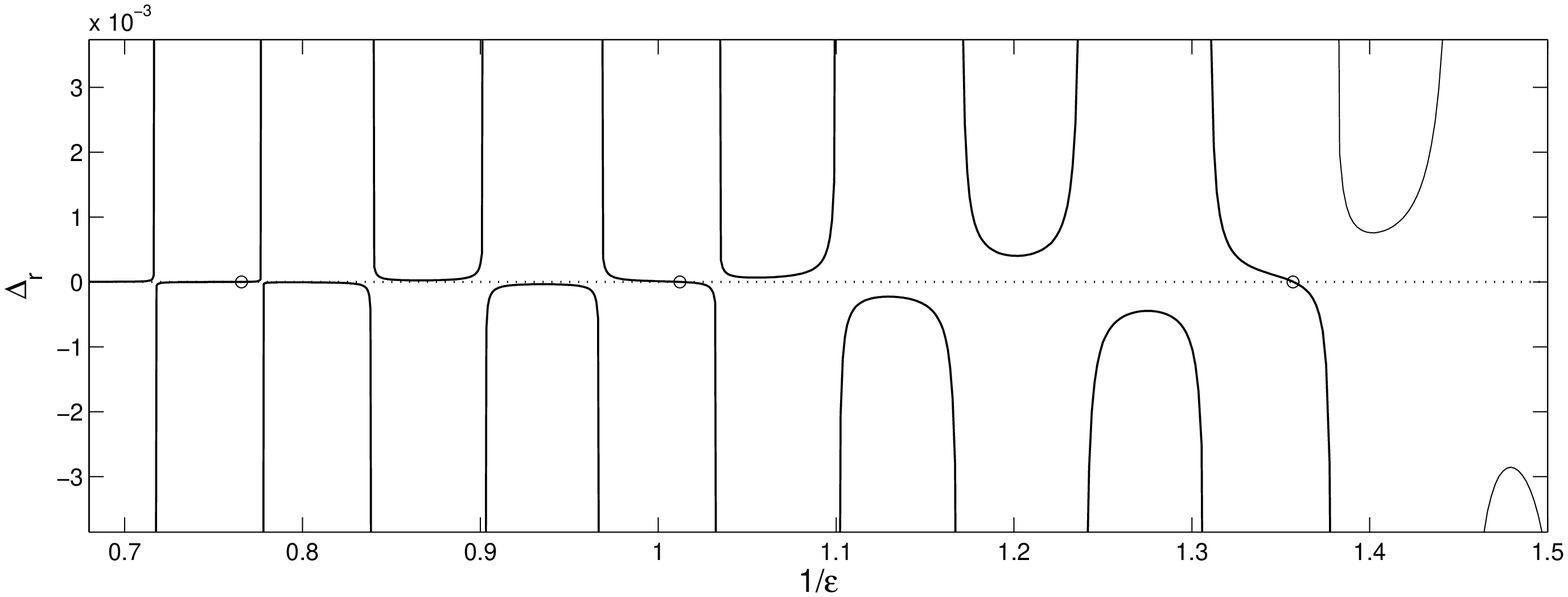}}
\subfigure[] { \label{Echeckzeros1}
\includegraphics[width=13cm,clip=]{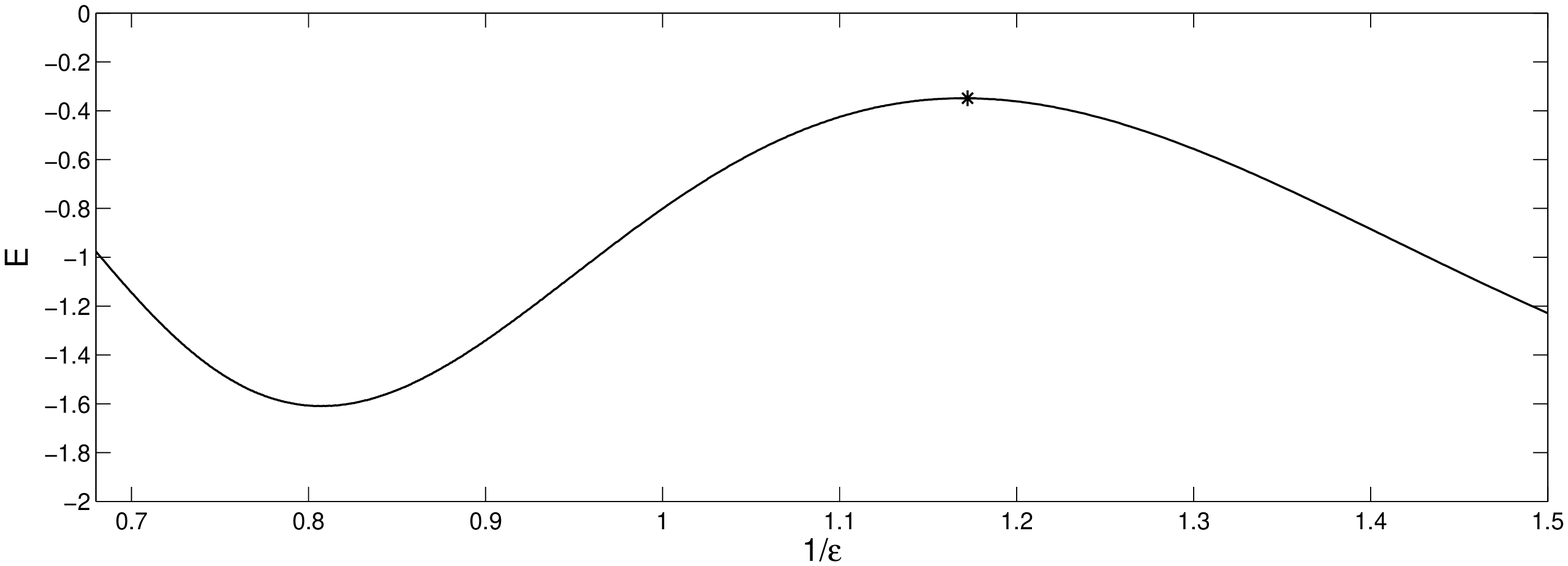}}
\caption{(a) The signed measure $\Delta _{r}$ (cf.~Eq.~(\protect\ref%
{measure1})) as a function of $1/\protect\varepsilon $ for $c=0.7$, $\Lambda
=0.5$, $\Delta z=0.2$, and $L=50$, where $\Delta _{r}$ is zero at $\protect%
\varepsilon \approx 0.737,0.988,1.306$ (or at $1/\protect\varepsilon \approx
1.357,1.012,0.766$), as shown by empty circles. (b) $E$ versus $1/\protect%
\varepsilon $ (cf.~Eq.~(\protect\ref{integralfinal})), for the same
parameter values $(\Lambda ,c)$ as in panel (a) and for the corresponding
solutions for $(A,a,p)$ and root(s) $\protect\lambda $ obtained from the
Eqs.~(\protect\ref{c1})-(\protect\ref{c3}) and Eq.~(\protect\ref{sin}),
respectively. It is clearly seen that $E\neq 0$ in the observed domain of $1/%
\protect\varepsilon $. However, we can conjecture that ESs are located near
the maximum of $E$, i.e., at $\protect\varepsilon \approx 0.853$ (or at $1/%
\protect\varepsilon \approx 1.172$), as indicated by the star.}
\label{ES1}
\end{figure}

\begin{figure}[tbph]
\centering
\subfigure[] { \label{zerocrossmeasure2}
\includegraphics[width=13cm,clip=]{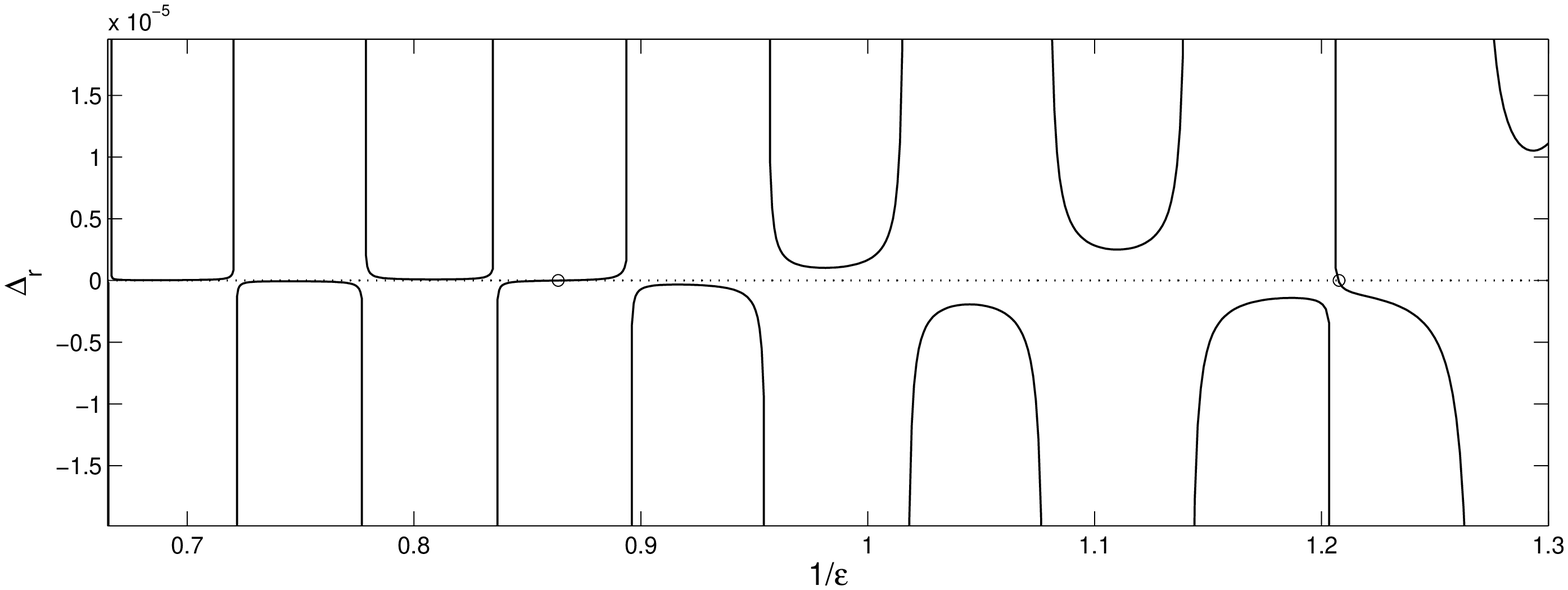}}
\subfigure[] { \label{Echeckzeros2}
\includegraphics[width=13cm,clip=]{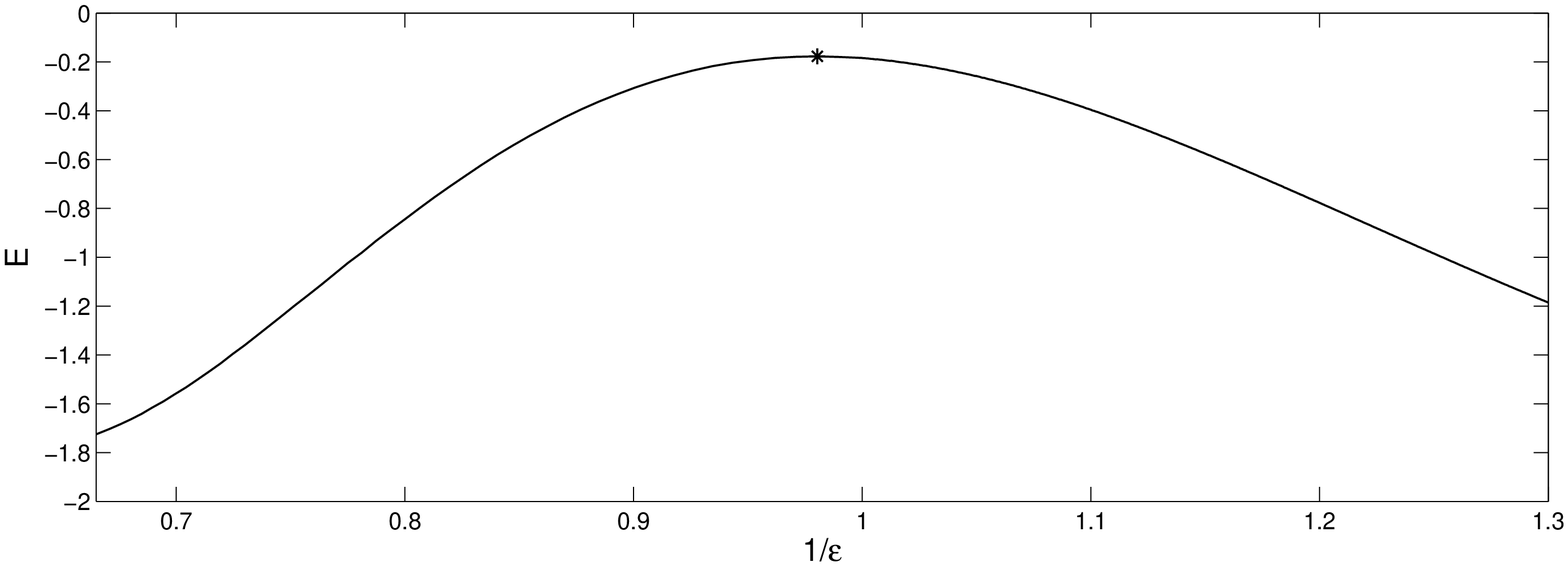}}
\caption{The same as Fig.~\protect\ref{ES1}, but for $\Lambda =0.7$. Zeros
of $\Delta _{r}$ are found at $\protect\varepsilon \approx 0.828,1.158$ (or
at $1/\protect\varepsilon \approx 1.2077,0.8636$), and the maximum of $E$
occurs at $\protect\varepsilon \approx 1.020$ (or at $1/\protect\varepsilon %
\approx 0.980$).}
\label{ES2}
\end{figure}

As shown in Ref.~\cite{Melvin}, in the general case the numerically obtained
solitary waves are weakly delocalized, i.e.,\ oscillatory tails with a
nonvanishing amplitude are attached to them. Nevertheless, as suggested in
Refs.~\cite{Melvin2,Melvin}, it is possible to find solutions with vanishing
tails (i.e., genuine solitons) by considering quantity
\begin{equation}
\Delta _{i}=\mathrm{Im}(\psi (N)),  \label{measure0}
\end{equation}%
which is a signed measure, whose zeros correspond to solutions with
vanishing tails. In the present case, the periodic boundary conditions lead
to $\mathrm{Im}(\psi (N))=0$, since the imaginary part of $\psi $ is odd.
Therefore, we modify the signed measure (\ref{measure0}) and redefine it as
\begin{equation}
\Delta _{r}=\mathrm{Re}(\psi (N)).  \label{measure1}
\end{equation}


In Fig.~\ref{zerocrossmeasure1} $\Delta _{r}$ is plotted as a function of $%
1/\varepsilon $ for the parameter values $(\Lambda ,c,\Delta z,L)$ used in
Fig.~\ref{AapVAxNum1}. We see from the figure that zeros of $\Delta _{r}$
occur at $\varepsilon \approx 0.737,0.988,1.306$ (i.e., at $1/\varepsilon
\approx 1.357,1.012,0.766$). For the parameter values in Fig.~\ref%
{AapVAxNum2}, we show the corresponding $\Delta _{r}$ in Fig.~\ref%
{zerocrossmeasure2}, from which we conclude that $\Delta _{r}$ vanishes at $%
\varepsilon \approx 0.828,1.158$ (i.e., at $1/\varepsilon \approx
1.208,0.864 $).
It is worthy of note that the plot of $\Delta _{r}$ also has singularities
which occur at exactly the same points as those
in Fig.~\ref{Aap} obtained from the soliton's core.

To predict the location of the ESs, we substitute the solution of Eqs.~(\ref%
{c1})--(\ref{c3}) and the root(s) $\lambda $ of Eq.~(\ref{sin}) into Eqs.~(%
\ref{integralfinal}) to find $E$ as a function of $\varepsilon ,c$, and $%
\Lambda $. Therefore, the existence of the ESs can be predicted by seeking
for values of the parameters at which $E=0$. For the parameter values used
in Figs.~\ref{zerocrossmeasure1} and \ref{zerocrossmeasure2},
curves for $E$ are displayed, respectively, in Figs. \ref{Echeckzeros1} and~%
\ref{Echeckzeros2}. 
It is seen that $E\neq 0$ in all the figures, i.e., truly localized solitons
cannot be directly predicted by the VA.

However, we can propose a conjecture, based on a \textquotedblleft
phenomenological\textquotedblright\ consideration of the figures, that there
are two zeros of $\Delta _{r}$ on the left and right of a maximum of $E$.
For example, in Fig.~\ref{ES1} we have the maximum of $E$ at $\varepsilon
\approx 0.853$ (i.e., at $1/\varepsilon \approx 1.172$), which is located
between two adjacent numerically found zeros of $\Delta _{r}$. The same
phenomenon also takes place in Fig.~\ref{ES2}
, where the two zero-crossing points lie between maxima of $E$, i.e., at $%
\varepsilon \approx 1.020$ (i.e., at $1/\varepsilon \approx 0.980$) in Fig.~%
\ref{ES2}. 
We have also computed (not shown here) the signed measure $\Delta _{r}$ and $%
E$ for other combinations of parameter values, where we observed the same
pattern. Thus, we conclude that, with addition of a constant, function $E$
may be able to predict the location of the ESs. We suspect that the missing
constant, which amounts to the shift of the plot for $E$ vertically, is
related to the choice of the ansatz (see, e.g., \cite{Boyd} for different
\textit{ans\"{a}tze} accounting for the oscillating tails).

%


\subsection{Stability}

To determine the VA-predicted stability of the soliton, we have solved the
eigenvalue problem (\ref{evpVA}). For the soliton shown in Fig.~\ref%
{comparedprofile1}, i.e., with $\mathbf{x}_{0}\approx \left[
1.228,0.554,0.377,0,0,0\right] ^{T}$, we obtain the corresponding
eigenvalues $\Omega \approx 0,0,0,0,0.436i,-0.436i$. In addition, for the
soliton in Fig.~\ref{comparedprofile2}, i.e., $\mathbf{x}_{0}\approx \left[
0.685,0.414,0.368,0,0,0\right] ^{T}$, the corresponding eigenvalues are $%
\Omega \approx 0,0,0,0,0.225i,-0.225i$. As the real part of all eigenvalues
is zero, we conclude that both solitons are \emph{stable}. These results are
in agreement with the numerical findings of Ref.~\cite{Melvin}.

\subsection{Bound states}

\begin{figure}[tbph]
\centering
\subfigure[$|l|\approx39.6$] { \label{bsL200e07l40}
  \includegraphics[width=16cm,clip=]{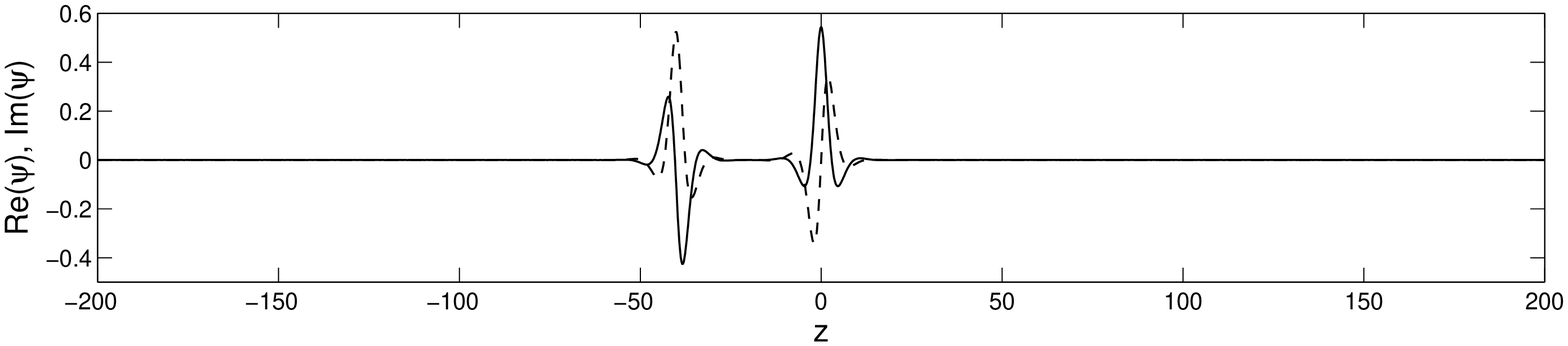}
  } \hspace{0cm}
\subfigure[$|l|\approx60.87$] { \label{bsL200e07l61}
  \includegraphics[width=16cm,clip=]{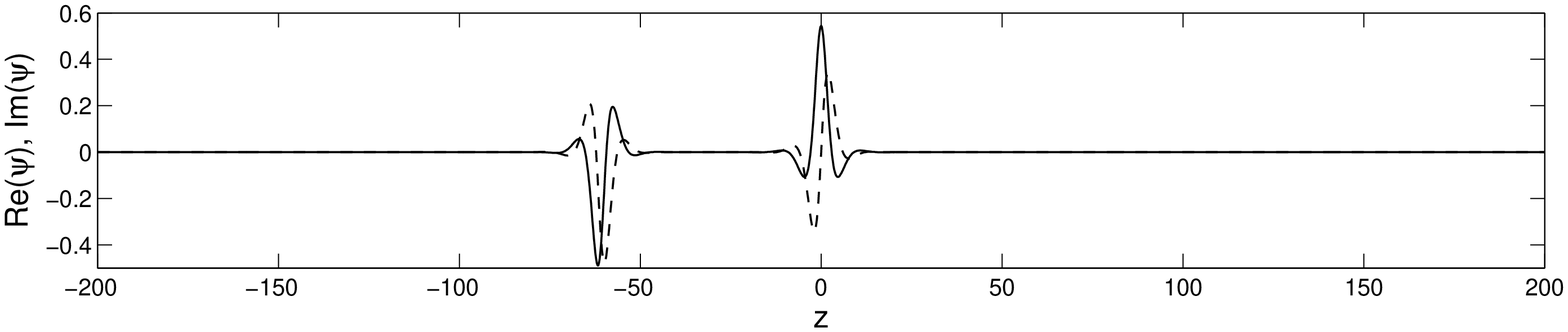}
  }
\caption{Two bound states, for $\protect\varepsilon =0.7$, $c=0.7$, $\Lambda
=0.7$, and $\Delta z=0.2$, found numerically in the interval of $[-200,200]$%
, for different distances between the two humps, $|l|$, as indicated in each
panel. Solid and dashed lines depict $\text{Re}(\protect\psi )$ and $\text{Im%
}(\protect\psi )$, respectively.}
\label{boundstate}
\end{figure}

Two examples of bound states consisting of two solitons, found numerically,
are presented in Fig.~\ref{boundstate}, each for the same parameter values.

Next, we compare the prediction presented above with the numerical results
shown in Fig.~\ref{boundstate}. For the parameter values used in that
figure, we have $\lambda _{r}\approx 0.441$ and $\lambda _{i}\approx 0.505$.
Note that, in terms of the above analysis, the soliton centred at $z=0$ in
Fig.~\ref{boundstate} is in fact soliton No.~2. Therefore, phase difference $%
\phi $ in this case is calculated as the phase of the soliton on the right
minus the phase of its counterpart on the left. From the numerical results,
the phase difference between the two solitons shown in the top and the
bottom panels of Fig.~\ref{boundstate} is given, respectively, by $\phi
\approx -2.012~$and $\phi \approx 2.368,$ both of which correspond to
$c\lambda _{i}/\lambda _{r}>0$, hence $n\cdot \mathrm{sign}\left( \lambda
_{i}\right) $ is odd. Using Eq.~(\ref{ln}), we find that the best fit to the
numerically computed distance in Fig.~\ref{boundstate}, i.e.,~$|l|\approx
39.6$ and $|l|\approx 60.87$, is provided by $l_{7}\approx 41.252$ and $%
l_{9}\approx 62.367$. Considering the fact that Eq.~(\ref{ln}) is based on a
simple approximation (Eqs.~(\ref{psi0}), (\ref{bckg1}) and (\ref{bckg2}))
and the assumption that the interacting solitons have decaying oscillating
tails, the approximation is in relatively good accordance with the results
provided by the numerical solution of the full equation.

\begin{figure}[tbph]
\centering
\subfigure[$l\approx39.6$] { \label{numevolbse07L200l40}
  \includegraphics[width=7.5cm,clip=]{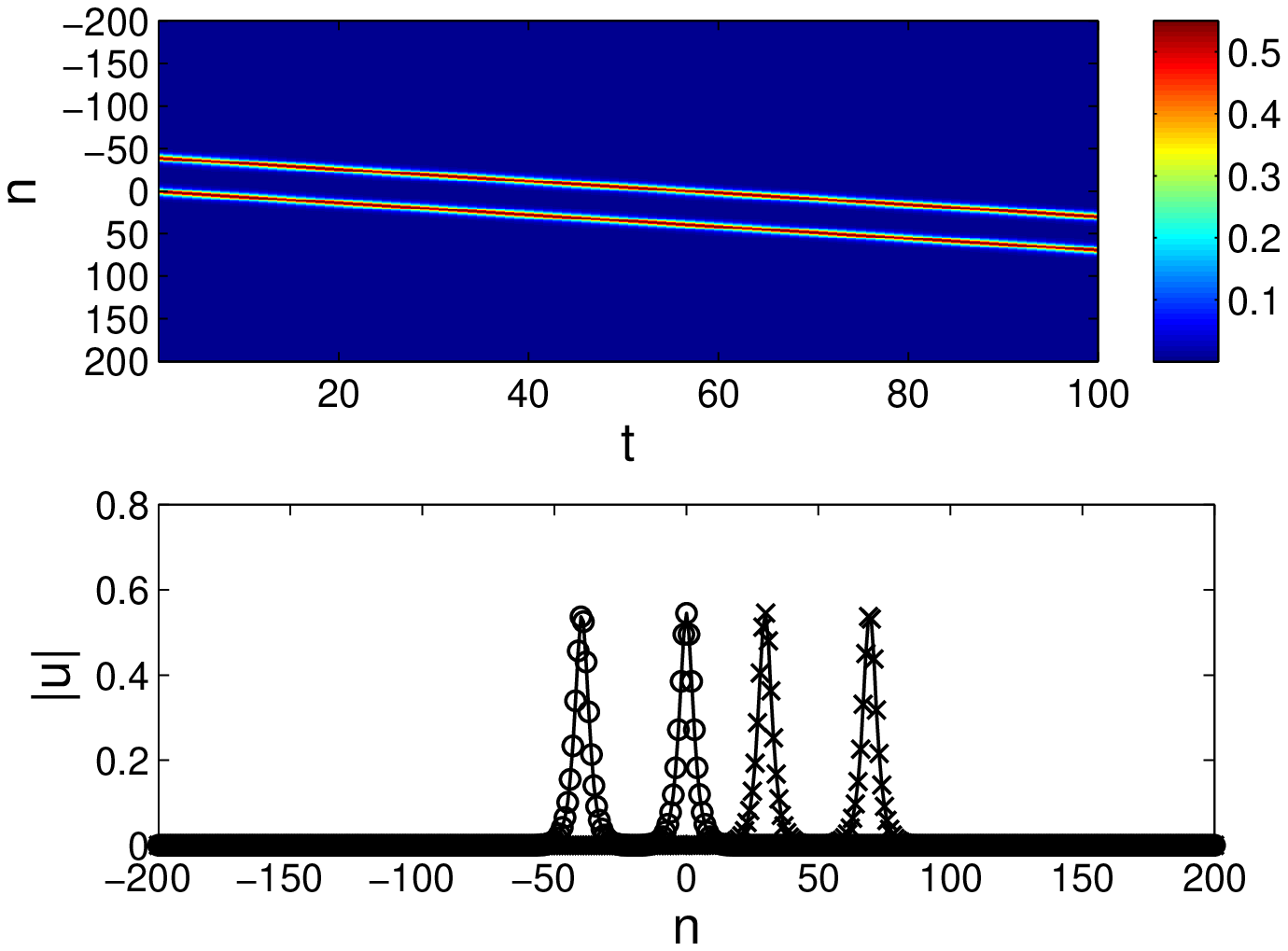}
  } \hspace{0cm}
\subfigure[$l\approx60.87$] { \label{numevolbse07L200l61}
  \includegraphics[width=7.5cm,clip=]{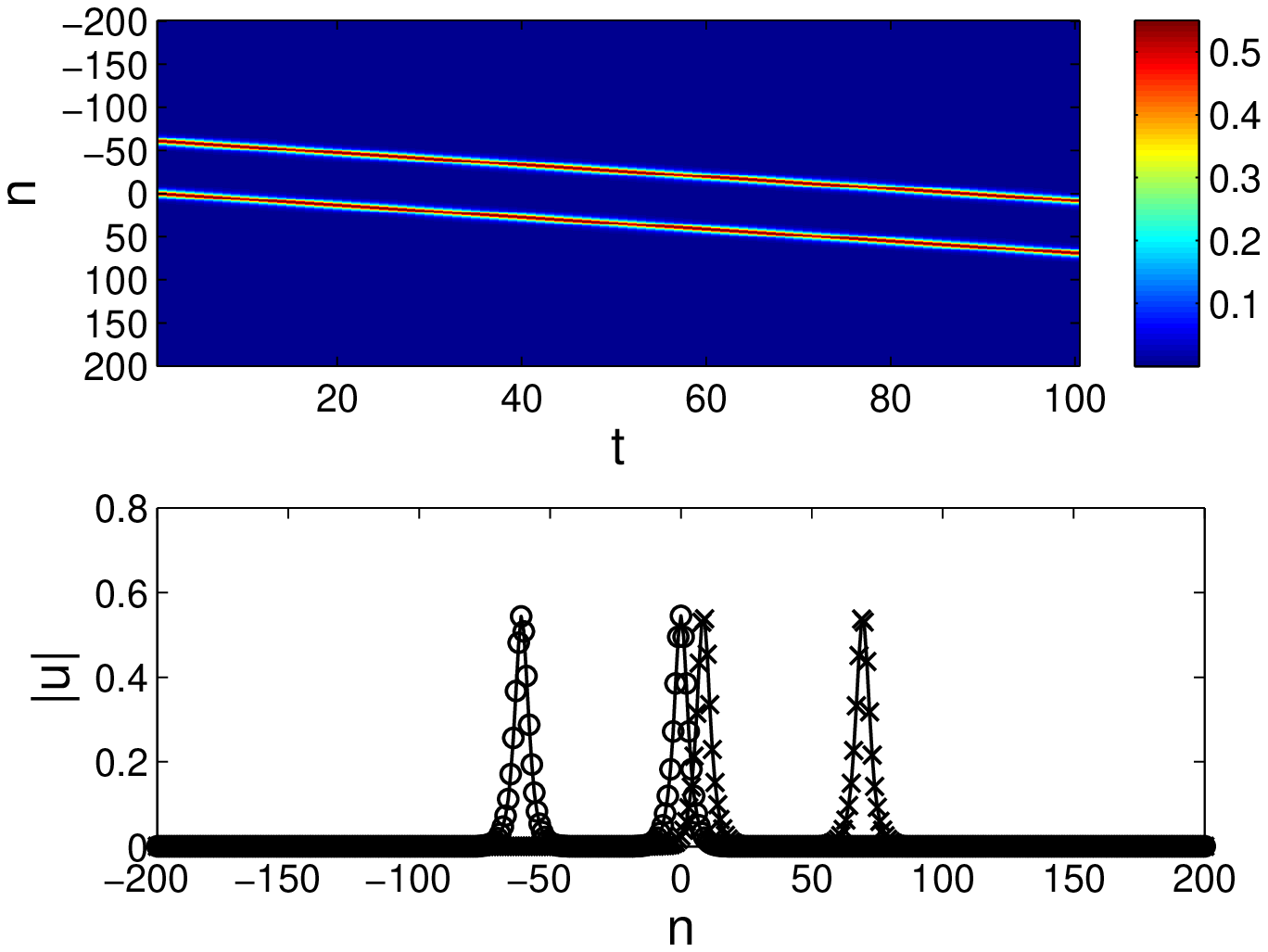}
  }
\caption{Numerical evolution of the solutions presented in Fig.~\protect\ref%
{boundstate}, obtained by means of a numerical grid with 401 sites. Upper
figures in each panel display the motion of the solution through the lattice
for the first 100 time units. Shown is a spatiotemporal contour plot of the
absolute value of the solution. Bottom figures in each panel depict the
initial (open circles) and final (crosses) profiles of the absolute value of
the solution after 100 time units of the evolution.}
\label{numevolbs}
\end{figure}

Using the Runge--Kutta method, we have also carried out numerical
integration of evolution equation (\ref{DNLSph}), with initial conditions
taken as in Fig.~\ref{boundstate}. The resulting evolution of the solutions
is displayed in Fig.~\ref{numevolbs}, where we see that both solitons
maintain their shapes and positions for a relatively long time.

\section{Conclusion}

The aim of this work is to develop the semi-analytical approach to seeking
travelling solitons, based on the application of the VA to the
differential--difference form of the DNLSE with the saturable nonlinearity,
in the moving coordinate frame. The predicted shapes of the solitons are in
good agreement with the numerical findings. The VA is also extended to
examine the stability of the travelling solitons, showing that they are
stable, which is consistent with previous work \cite{Melvin}. Further, the VA
was developed to predict the locations of the exceptional solutions for
genuine travelling solitons with strictly vanishing tails. Bound states of
two solitons were briefly considered too. In the latter case, the VA
predicts the distance between two solitons forming the bound state. In the
numerical part of the work, we have made use of the numerical scheme for the
DNLSE written in the moving coordinate frame, which is an equation of the
differential--difference type. Using the Newton--Raphson method, we have
confirmed the existence of exceptional solutions for travelling discrete
solitons (which are \textquotedblleft embedded solitons", in this sense),
earlier predicted by means of a different numerical algorithm. We have
compared the analytical results based on the VA and the numerical findings,
concluding that they are in good agreement.

\section*{Acknowledgement}

MS acknowledges the Ministry of National Education of the Republic of
Indonesia for financial support.



%



\end{document}